\documentclass[authoryear,3p,times,final]{elsarticle}

\usepackage{amssymb,multicol}
\usepackage{amsmath,comment}

\usepackage{graphicx,algorithm,epstopdf,lineno}

\usepackage{color,url,float,epsf,subfig,tabularx}
\usepackage{microtype,bm}
\usepackage[colorlinks=true,breaklinks=true,pdftex]{hyperref}

\newtheorem{Theorem}{Theorem}
\newdefinition{definition}{Definition}
\newdefinition{Remark}{Remark}
\newdefinition{cor}{Corollary}
\newdefinition{claim}{Claim}
\newdefinition{proposition}{Proposition}

\newdefinition{Example}{Example}

\journal{arXiv.org}
\date{}
\begin{document}

\begin{frontmatter}

\title{Efficient Subdivision of B\'{e}zier Curves/Surfaces via Blossoms}

\author[inst1]{Krassimira Vlachkova}
\cortext[mycorrespondingauthor]{Corresponding author}
\ead{krassivl@fmi.uni-sofia.bg}

\affiliation[inst1]{organization={Faculty of Mathematics and Informatics, Sofia University ``St. Kliment Ohridski"},
	            addressline={Blvd. James Bourchier 5},
                city={Sofia},
	            postcode={1164},
	            country={Bulgaria}}

\begin{abstract}
We consider the problem of B\'{e}zier curves/surfaces subdivision using blossoms. We propose closed-form formulae for blossoms evaluation, as needed for the calculation of control points. This approach leads to direct and efficient way to obtain subdivisions  for B\'{e}zier curves and both tensor product and triangular B\'{e}zier surfaces. 
It  simplifies considerably the computation of control points of subdivisions which is crucial in applications where curves/surfaces need to be refined or adapted dynamically. For instance, in CAD/CAM systems, architectural design, or animation, the ability to quickly and accurately determine new control points is essential for manipulation and rendering complex shapes. More efficient subdivision can facilitate complex operations like finding intersections between surfaces or smoothly blending multiple surfaces.
\end{abstract}

\begin{keyword}
	B\'{e}zier curve \sep tensor product B\'{e}zier surface \sep triangular B\'{e}zier surface \sep blossom \sep subdivision
\MSC 65D17 \sep 65D18 \sep 65D19 \sep 68U05 \sep 68U07
\end{keyword}
\pagebreak
\end{frontmatter}


\section{Introduction}

Blossoming is a fundamental tool for analyzing polynomial and piecewise polynomial curves and surfaces. It enables derivation of important identities and algorithms \citep{RG,M}. It was originally introduced to the field of Computer Aided Geometric Design (CAGD) by de Casteljau \citep{deCasteljau} and later developed by Ramshaw  \citep{ramshaw1987,ramshaw1989} for the study of B\'{e}zier and B-splines curves and surfaces.

The blossom of a polynomial $P(t)$ of degree $n$ is the unique multivariate polynomial $p=p(u_1,\dots ,u_n)$ satisfying the following properties \citep{RG}:
\begin{enumerate}
	\item[(i)] {\it  Symmetry.} Polynomial $p$ is invariant under permutation of its parameters, i.e.
	$p(u_{\sigma (1)},\dots , u_{\sigma (n)}) = p(u_1, \dots, u_n)$ for any permutation $\sigma$ of $\{1, \dots , n\}$.
	\item[(ii)] {\it Multi-affinity.} Polynomial $p$ is affine in each of its variables
		i.e.\\ $p(\dots ,(1-\alpha)u_k+\alpha v_k,\dots)=(1-\alpha)p(\dots ,u_k,\dots)+\alpha p(\dots ,v_k,\dots).$
	\item[(iii)] {\it Diagonal reduction.} Polynomial $p$ reduces to the original polynomial $P(t)$ along the diagonal, i.e. $p(t,\dots ,t)=P(t).$
\end{enumerate}
The definition of blossom in the bivariate case is extended straightforward \citep{RG}. 

Blossoming approach offers significant utility in various aspects of CAGD theory including curve and surface manipulation, 
computing derivatives and identities, establishing  smoothness conditions, changing bases, dual functionals and others, see \citep{RG,M}. 
In CAD systems blossoming facilitates evaluation and manipulation of control polygons and nets for subdivisions, enables high precision geometric representations of complex surfaces, supports smooth animation via rapid computation and interpolation of control points, enhances control over surface continuity and adaptability, enabling the design of intricate shapes in various engineering fields and others.

Beyond polynomials, the concept of blossoming has been generalized to various function spaces and systems, including quantum Bernstein bases  \citep{goldman2015quantum}, Chebyshev systems  \citep{mazure2006ready}, trigonometric splines  \citep{gonsor1996null}, and R-blossoming  \citep{goldman2025universal}, which encompasses many of these as special cases.
While the core concepts of blossoming were implicitly present in earlier work \citep{A-HM2}, recent progress emphasizes extending the blossoming beyond its traditional polynomial framework to a more unified and universal approach that extends blossoming to all finite-dimensional spaces of univariate functions
\citep{goldman2025universal}. Recent work also highlights significant progress in understanding and constructing complex curves, introducing novel functional concepts like the multirational blossom \citep{TSG2,TSG1}, and developing improved approximation and interpolation techniques \citep{A-HM1,BEL,F-J,BEIPL,SEHC,UWQD,LL}. These advancements contribute to better accuracy and deeper theoretical insight within their respective fields.

In this paper we consider the following three problems: 
\begin{enumerate}
	\item[(1)]  Given a polynomial curve  $C(u)=\sum_{i=0}^n{\bf c}_{i}u^i$ of degree $n$, $n\in\mathbb{N}$, ${\bf c}_{i}\in\mathbb{R}^3$, $i=0,\dots ,n$, find B\'{e}zier curve $C^1$ on $C$ defined for  $a\leq u\leq b$, where $a,b\in\mathbb{R}$;
	\item[(2)] Given a polynomial surface  $S(u,v)=\sum_{i=0}^n\sum_{j=0}^m{\bf c}_{ij}u^iv^j$ of degree $(n,m)$, $n,m\in\mathbb{N}$, ${\bf c}_{ij}\in\mathbb{R}^3$, $i=0,\dots ,n$, $j=0,\dots ,m$,  find tensor product B\'{e}zier (TPB) surface patch $S^1$ on $S$ defined for  $a\leq u\leq b$ and $c\leq v\leq d$, where $a,b,c,d\in\mathbb{R}$;
	\item[(3)] Given a polynomial surface  $S(u,v)=\sum_{i=0}^n\sum_{j=0}^m{\bf c}_{ij}u^iv^j$ of degree $(n,m)$, $n,m\in\mathbb{R}^3$, ${\bf c}_{ij}\in\mathbb{N}$, $i=0,\dots ,n$, $j=0,\dots ,m$,  find triangular B\'{e}zier (TB) surface patch $S^2$ on $S$ defined in a triangle  $\triangle \bf{abc}$, where ${\bf a},{\bf b},{\bf c}\in\mathbb{R}^2$.
\end{enumerate}

Standard approach to derive the B\'{e}zier curve $C^1$ and the B\'{e}zier surface patches $S^1$ and $S^2$ is subdivision using blossoming as it is detailed in Section~\ref{sect2}.
The control points of $C^1$, $S^1$, and $S^2$ are expressed in terms of the values of the corresponding blossoms. They solely depend on the parameters $a,b$; $a,b,c,d$; $\bf{a},\bf{b},\bf{c}$, respectively. However, the relevant expressions for the control points are long and computationally intractable sums.
We propose a closed-form formulae for computation of the control points for each of the above three problems. These formulae provide a straightforward and effective procedure for computation of control points and facilitate the evaluation process for the three considered problems.

The paper is organized as follows: In Section~\ref{sect2} we describe the solutions of the considered problems by blossoming. Then  we present closed-form formulae for computation of the control points and prove their correctness. In Section~\ref{sect3} we present and discuss results from the numerical implementation and visualization of the closed-form formulae.
At the end we provide some concluding remarks.

\section{Main Results}\label{sect2}
\subsection{Subdivision of a B\'{e}zier Curve}

Let $C(u)=\sum_{i=0}^n{\bf c}_{i}u^i$ be a polynomial curve of degree $n$, $n\in\mathbb{N}$, ${\bf c}_{i}\in\mathbb{R}^3$, $i=0,\dots ,n$.  Let the B\'{e}zier curve $C^1(u)=\sum_{\nu =0}^n{\bf w}_{\nu}B_{\nu}^n(u)$ be the part of $C(u)$ defined in the interval $[a,b]$, where $a,b\in\mathbb{R}$, ${\bf w}_{\nu}$ are the control points of $C^1$, and $B_{\nu}^n(u)=\binom{n}{\nu}u^{\nu}(1-u)^{n-\nu}$ are the Bernstein polynomials, $\nu =0,\dots ,n$.

Let $b\tilde{\ }(u_1,\dots ,u_n)$ be the blossom of $C(u)$. It is well known that
\begin{equation}\label{ec11}
	{\bf w}_{\nu}=b\tilde{\ } (\underbrace{b,\dots ,b}_\text{$\nu$},\underbrace{a,\dots ,a}_\text{$n-\nu$}).
\end{equation}
	Furthermore, for the blossom $b\tilde{\ }$ we have the formula
	$$b\tilde{\ }(u_1,\dots ,u_n)=\sum_{i=0}^n{\bf c}_{i}b_{i}\tilde{\ }(u_1,\dots ,u_n),\ {\rm where}$$
	
	\begin{equation}\label{ec12}	
		b_{i}\tilde{\ }(u_1,\dots ,u_n)=\sum_{\{\alpha_1,\dots ,\alpha_i\}\subseteq \{1,\dots ,n\}}\frac{u_{\alpha_1}\dots u_{\alpha_i}}{\binom{n}{i}}
	\end{equation}
	is the blossom of the monomial $u^i$. The sums in (\ref{ec12}) are taken over all distinct indices $\{\alpha_1,\dots \alpha_i\}$
	and $b_{0}\tilde{\ }(u_1,\dots ,u_n)=1$.

	We have observed that the evaluation of blossoms using formula (\ref{ec12}) is not straightforward since it requires the generation of all subsets
	$\{\alpha_1,\dots ,\alpha_i\}$. This make the evaluation of control points computationally intense. In the next theorem we propose a formula for direct computation of the control points ${\bf w}_{\nu}$, $\nu =0,\dots ,n$.

	\begin{Theorem}\label{th0}
		The B\'{e}zier control points ${\bf w}_{\nu}$, $\nu =0,\dots ,n$, defined by (\ref{ec11}) are
		\begin{equation}\label{eqc1}
			{\bf w}_{\nu}=
			\sum_{i=0}^n\frac{{\bf c}_{i}}{\binom{n}{i}}\sum_{k=\max (0,i+\nu-n)}^{\min (i,\nu)} \binom{\nu}{k}\binom{n-\nu}{i-k}
			b^ka^{i-k}.
		\end{equation}
		
		{\bf Proof} According to (\ref{ec12}), we have
			$$b_{i}\tilde{\ } (u_1,\dots ,u_n)
			:=\frac{1}{\binom{n}{i}}\sum_{\bar{\alpha}_i\in  {\cal A}_i}U(\bar{\alpha}_i)
			$$
			where $ {\cal A}_i$ is the set of all subsets $\bar{\alpha}_i:=\{\alpha_1,\dots ,\alpha_i\}$ of distinct indices of size $i$ and $U(\bar{\alpha}_i):=u_{\alpha_1}\dots u_{\alpha_i}$.
			
			Next, we will compute $b_{i}\tilde{\ } (\underbrace{b,\dots ,b}_\text{$\nu$},\underbrace{a,\dots ,a}_\text{$n-\nu$})$. For a fixed $\bar{\alpha}_i$ let $k=k(\bar{\alpha}_i)$ be the number of indices in $\bar{\alpha}_i$ that are in $[1,\nu]$. Then the number of indices in $\bar{\alpha}_i$ that are in $[\nu +1,n]$ is $i-k$. Note that the following inequalities must hold
			\begin{equation}\label{ec13}
				k\leq i,\ k\leq\nu,\ i-k\leq n-\nu .
			\end{equation}
			In this notation for a fixed $\bar{\alpha}_i$ we have $U(\bar{\alpha}_i)=b^ka^{i-k}$. Consequently,
			$$\sum_{\bar{\alpha}_i\in  {\cal A}_i}U(\bar{\alpha}_i)=\sum_{k=0}^iI(k)b^ka^{i-k}$$
			where $I(k)$ is the number of subsets $\bar{\alpha}_i$ that have $k$ indices in $[1,\nu]$. We have $I(k)=\binom{\nu}{k}\binom{n-\nu}{i-k}$ and hence, from (\ref{ec13}) it follows
			\begin{equation}\label{ec14}
				\sum_{\bar{\alpha}_i\in  {\cal A}_i}U(\bar{\alpha}_i)=\sum_{k=\max (0,i+\nu-n)}^{\min (i,\nu)}\binom{\nu}{k}\binom{n-\nu}{i-k}b^ka^{n-k}.
			\end{equation}
			Substituting (\ref{ec14}) in (\ref{ec12}) and (\ref{ec11}), we get
			$${\bf w}_{\nu}=
			\sum_{i=0}^n\frac{{\bf c}_{i}}{\binom{n}{i}}\sum_{k=\max (0,i+\nu-n)}^{\min (i,\nu)}\binom{\nu}{k}\binom{n-\nu}{i-k}
			b^ka^{i-k}. $$
	\hfill$\Box$

	\end{Theorem}

	\subsection{Subdivision of a Tensor Product B\'{e}zier Surface }
	
	Let $S(u,v)=\sum_{i=0}^n\sum_{j=0}^m{\bf c}_{ij}u^iv^j$ be a polynomial surface of bidegree $(n,m)$, $n,m\in\mathbb{N}$, ${\bf c}_{ij}\in\mathbb{R}^3$, $i=0,\dots ,n$, $j=0,\dots ,m$. Let the TPB surface $S^1(u,v)=\sum_{\nu =0}^n\sum_{\mu =0}^m{\bf p}_{\nu\mu}B_{\nu}^n(u)B_{\mu}^n(v)$ be the part of $S(u,v)$ defined for $a\leq u\leq b$ and $c\leq v\leq d$, where $a,b,c,d\in\mathbb{R}$ and ${\bf p}_{\nu\mu}$ are the control points of $S^1$, $\nu =0,\dots ,n$, $\mu =0,\dots ,m$.
	
	Let $b^{\Box}(u_1,\dots ,u_n,v_1,\dots ,v_m)$ be the blossom of $S(u,v)$. It is known that
	\begin{equation}\label{e11}
		{\bf p}_{\nu\mu}=b^{\Box} (\underbrace{b,\dots ,b}_\text{$\nu$},\underbrace{a,\dots ,a}_\text{$n-\nu$},\underbrace{d,\dots ,d}_\text{$\mu$},\underbrace{c,\dots ,c}_\text{$m-\mu$}).
	\end{equation}
	Furthermore, for the blossom $b^{\Box}$ we have the formula (see \citep{RG}, pp. 338-339)
	\begin{equation}\label{ecor}
		b^{\Box}(u_1,\dots ,u_n,v_1,\dots ,v_m)=\sum_{i=0}^n\sum_{j=0}^m{\bf c}_{ij}b_{ij}^{\Box} (u_1,\dots ,u_n,v_1,\dots ,v_m),\ \rm{where}
	\end{equation}
	\begin{equation}\label{e12}
		b_{ij}^{\Box} (u_1,\dots ,u_n,v_1,\dots ,v_m)
		=\sum_{\{\alpha_1,\dots ,\alpha_i\}\subseteq \{1,\dots ,n\}}\frac{u_{\alpha_1}\dots u_{\alpha_i}}{\binom{n}{i}} \sum_{\{\beta_1,\dots ,\beta_j\}\subseteq \{1,\dots ,m\}}\frac{v_{\beta_1}\dots v_{\beta_j}}{\binom{m}{j}}
	\end{equation}
	is the blossom  of the monomial $u^iv^j$. The sums in (\ref{e12}) are taken over all distinct indices $\{\alpha_1,\dots \alpha_i\}$ and all distinct indices $\{\beta_1,\dots \beta_i\}$. In the case where either $i=0$, or $j=0$, the corresponding sum in (\ref{e12}) equals 1. 
	
	We note that formula (\ref{e12}) in this form is not easy to be  implemented. Therefore, next we establish formula (\ref{e11})
	in an equivalent closed form that is more suitable for computations.

	\begin{Theorem}\label{th1}
		The B\'{e}zier control points ${\bf p}_{\nu\mu}$, $\nu=0,\dots ,n$, $\mu=0,\dots ,m$, defined by (\ref{e11}) are
		\begin{equation}\label{eq1}
			{\bf p}_{\nu\mu}=
			\sum_{i=0}^n\sum_{j=0}^m\frac{{\bf c}_{ij}}{\binom{n}{i}\binom{m}{j}}\sum_{k=\max (0,i+\nu-n)}^{\min (i,\nu)}\ \sum_{r=\max (0,j+\mu-m)}^{\min (j,\mu)} \binom{\nu}{k}\binom{n-\nu}{i-k}\binom{\mu}{r}\binom{m-\mu}{j-r}
			b^ka^{i-k}d^rc^{j-r}.
		\end{equation}
		
	\end{Theorem}
	{\bf Proof} We observe that formula (\ref{e12}) can be presented in the   equivalent form 
		\begin{equation}\label{bloss1}
			b_{ij}^{\Box} (u_1,\dots ,u_n,v_1,\dots ,v_m)=b_{i}\tilde{\ } (u_1,\dots ,u_n) b_{j}\tilde{\ } (v_1,\dots ,v_m), 
		\end{equation}
		where	$	b_{i}\tilde{\ }$ and $	b_{j}\tilde{\ }$ are the blossoms of the monomials $u^i$ and $v^j$, respectively.
		From Theorem~\ref{th0} we have
		\begin{eqnarray}
			&&	b_{i}\tilde{\ } (u_1,\dots ,u_n)=	\frac{1}{\binom{n}{i}}\sum_{k=\max (0,i+\nu-n)}^{\min (i,\nu)}\binom{\nu}{k}\binom{n-\nu}{i-k}b^ka^{n-k},\nonumber\\
			&&\label{short}\\
			&&b_j\tilde{\ } (v_1,\dots ,v_m)=\frac{1}{\binom{m}{j}}\sum_{r=\max (0,j+\mu-m)}^{\min (j,\mu)}\binom{\mu}{r}\binom{m-\mu}{j-r}
			d^rc^{j-r}.\nonumber
		\end{eqnarray}	
		Substituting (\ref{short})  in (\ref{bloss1}) and (\ref{ecor}),  we obtain from (\ref{e11})
		$${\bf p}_{\nu\mu}=
		\sum_{i=0}^n\sum_{j=0}^m\frac{{\bf c}_{ij}}{\binom{n}{i}\binom{m}{j}}\sum_{k=\max (0,i+\nu-n)}^{\min (i,\nu)}\ \sum_{r=\max (0,j+\mu-m)}^{\min (j,\mu)} \binom{\nu}{k}\binom{n-\nu}{i-k}\binom{\mu}{r}\binom{m-\mu}{j-r}
		b^ka^{i-k}d^rc^{j-r}. $$
	\hfill$\Box$

	\subsection{Subdivision of a Triangular B\'{e}zier Surface }
	
	Let $S(u,v)=\sum_{i=0}^n\sum_{j=0}^m{\bf c}_{ij}u^iv^j$ be a polynomial surface of total degree $N=n+m$, $n,m\in\mathbb{N}$, ${\bf c}_{ij}\in\mathbb{R}^3$, $i=0,\dots ,n$, $j=0,\dots ,m$, and ${\bf a}=(a_1,a_2)$, ${\bf b}=(b_1,b_2)$, ${\bf c}=(c_1,c_2)$ be three non-collinear points in $\mathbb{R}^2$. Let the TB surface
	$$S^2(u,v)=\sum_{\nu =0}^N\sum_{\mu =0}^{N-\nu}{\bf q}_{\nu\mu}B_{\nu\mu}^N(u,v)u^{\nu}v^{\mu}(1-u-v)^{N-\nu-\mu}$$
	be the part of $S(u,v)$ defined
	in the triangle $\triangle \bf{abc}$, where ${\bf q}_{\nu\mu}$ are the control points of $S^2$,
	$$B_{\nu\mu}^N(u,v)=\binom{N}{\nu ,\mu}u^{\nu}v^{\mu}(1-u-v)^{N-\nu-\mu},\ \nu=0,\dots ,N,\ \mu =0,\dots ,N-\nu $$
	are the bivariate Bernstein polynomials, and $\binom{N}{\nu,\mu}=\frac{N!}{\nu !\mu !(N-\nu-\mu)!}$.
	
	Let $b^{\triangle}((u_1,v_1),\dots ,(u_{N},v_{N}))$ be the blossom of $S(u,v)$. It is known that
	
	\begin{equation}\label{e31}
		{\bf q}_{\nu\mu}=b^{\triangle} (\underbrace{{\bf a},\dots ,{\bf a}}_{\nu},\underbrace{{\bf b},\dots ,{\bf b}}_{\mu},\underbrace{{\bf c},\dots {\bf c}}_{N-\nu-\mu}).
	\end{equation}
	Furthermore, for the blossom $b^{\triangle}$ we have the formula (see \citep{RG}, pp. 338-339)
	$$b^{\triangle}((u_1,v_1),\dots ,(u_{N},v_{N}))=\sum_{i=0}^n\sum_{j=0}^m{\bf c}_{ij}b_{ij}^{\triangle}((u_1,v_1),\dots ,(u_{N},v_{N})),\ \rm{where}$$
	\begin{equation}\label{e32}
		b_{ij}^{\triangle} ((u_1,v_1),\dots ,(u_{N},v_{N}))=\sum_{\scriptstyle\{\alpha_1,\dots ,\alpha_i,\beta_1,\dots ,\beta_j\}\subseteq \{1,\dots, N\}}
		\frac{u_{\alpha_1}\dots u_{\alpha_i}v_{\beta_1}\dots v_{\beta_j}}{\binom{N}{i,j}},
	\end{equation}
	is the blossom  of the monomial $u^iv^j$. The sum in (\ref{e32}) is taken over all sets of distinct indices $\{\alpha_1,\dots ,\alpha_i,\beta_1,\dots ,\beta_j\}$ and
	$b_{i0}^{\triangle} ((u_1,v_1),\dots ,(u_{N},v_{N}))=
	b_i\tilde{\ }(u_1,\dots ,u_N)$, $b_{0j}^{\triangle} ((u_1,v_1),\dots ,(u_{N},v_{N}))=
	b_j\tilde{\ }(v_1,\dots ,v_N).$
	
	We note that the evaluation of blossoms using formula (\ref{e32}) is not straightforward and easy for  implementation. Therefore, similar to the case of TPB surfaces, in the next theorem we propose a formula for direct computation of the control points ${\bf q}_{\nu\mu}$, $\nu =0,\dots ,N$, $\mu =0,\dots ,N-\nu$.

	\begin{Theorem}\label{th2}
		The B\'{e}zier control points ${\bf q}_{\nu\mu}$, $\nu=0,\dots ,N$, $\mu=0,\dots ,N-\nu$, defined by (\ref{e31}) are
		
		\begin{eqnarray}	
			{\bf q}_{\nu\mu}&=&\sum_{i=0}^n\sum_{j=0}^m\frac{{\bf c}_{ij}}{\binom{n+m}{i,j}}
			\sum_{i_{\alpha}=\underline{i}_{\alpha}}^{\bar{i}_{\alpha}}	
			\sum_{i_{\beta}=\underline{i}_{\beta}}^{\bar{i}_{\beta}}	
			\sum_{j_{\alpha}=\underline{j}_{\alpha}}^{\bar{j}_{\alpha}}
			\sum_{j_{\beta}=\underline{j}_{\beta}}^{\bar{j}_{\beta}}\nonumber		\\
			&&	\binom{\nu}{i_{\alpha}}
			\binom{\mu}{i_{\beta}}
			\binom{\lambda}{i_{\gamma}}
			\binom{\nu -i_{\alpha}}{j_{\alpha}}
			\binom{\mu -i_{\beta}}{j_{\beta}}
			\binom{\lambda -i_{\gamma}}{j_{\gamma}}
			a_1^{i_{\alpha}}b_1^{i_{\beta}}c_1^{i_{\gamma}}
			a_2^{j_{\alpha}}b_2^{j_{\beta}}c_2^{j_{\gamma}},\label{eq2}
		\end{eqnarray}
		
		where
		$$
		\begin{array}{l}
			\lambda=n+m-\nu -\mu ,\  i_{\gamma}=i-i_{\alpha}-i_{\beta},\ j_{\gamma}=j-j_{\alpha}-j_{\beta},\\
			\underline{i}_{\alpha}=\max (0,i-\mu -\lambda),\ \bar{i}_{\alpha}=\min (i,\nu),\\
			\underline{i}_{\beta}=\max (0,i-i_{\alpha}-\lambda),\ \bar{i}_{\beta}=\min (i-i_{\alpha},\mu),\\
			\underline{j}_{\alpha}=\max (0,j-(\mu-i_{\beta})-(\lambda - i_{\gamma})),\  \bar{j}_{\alpha}=\min (j,\nu -i_{\alpha}),\\
			\underline{j}_{\beta}=\max (0,j-j_{\alpha}-(\lambda - i_{\gamma})),\  \bar{j}_{\beta}=\min (j-j_{\alpha},\mu -i_{\beta}).
		\end{array}
		$$
	\end{Theorem}
	{\bf Proof.} First, we will compute $b_{ij}^{\triangle} (\underbrace{{\bf a},\dots ,{\bf a}}_{\nu},\underbrace{{\bf b},\dots ,{\bf b}}_{\mu},\underbrace{{\bf c},\dots {\bf c}}_{N-\nu-\mu})$. For fixed $\bar{\alpha}_i=\{\alpha_1,\dots, \alpha_i\}$ and $\bar{\beta}_j=\{\beta_1,\dots, \beta_j\}$ we have
		$$u_{\alpha_1}\dots u_{\alpha_i}v_{\beta_1}\dots v_{\beta_j}=a_1^{i_{\alpha}}b_1^{i_{\beta}}c_1^{i_{\gamma}}
		a_2^{j_{\alpha}}b_2^{j_{\beta}}c_2^{j_{\gamma}}\quad {\rm where} $$
		$$
		\begin{array}{l}
			i_{\alpha}\ {\rm is\ the\ number\ of\ indices\ in}\ \bar{\alpha}_i\ {\rm that\ are\ in}\ [1,\nu],\\
			i_{\beta}\ {\rm is\ the\ number\ of\ indices\ in}\ \bar{\alpha}_i\ {\rm that\ are\ in}\ [\nu+1,\nu+\mu],\\
			i_{\gamma}:=i-i_{\alpha}-i_{\beta}\ {\rm is\ the\ number\ of\ indices\ in}\ \bar{\alpha}_i\ {\rm that\ are\ in}\ [\nu+\mu +1,N],\\
			j_{\alpha}\ {\rm is\ the\ number\ of\ indices\ in}\ \bar{\beta}_j\ {\rm that\ are\ in}\ [1,\nu],\\
			j_{\beta}\ {\rm is\ the\ number\ of\ indices\ in}\ \bar{\beta}_j\ {\rm that\ are\ in}\ [\nu+1,\nu+\mu],\\	
			j_{\gamma}:=j-j_{\alpha}-j_{\beta}\ {\rm is\ the\ number\ of\ indices\ in}\ \bar{\beta}_j\ {\rm that\ are\ in}\ [\nu+\mu +1,N].
		\end{array}
		$$

		For a fixed quadruple $(i_{\alpha},i_{\beta},j_{\alpha},j_{\beta})$ let $I(i_{\alpha},i_{\beta},j_{\alpha},j_{\beta})$ be the set of index vectors $(\bar{\alpha}_i,\bar{\beta}_j)$ that have prescribed $i_{\alpha},i_{\beta},j_{\alpha},j_{\beta}$ numbers. Then we have
		$$b_{ij}^{\triangle} (\underbrace{{\bf a},\dots ,{\bf a}}_{\nu},\underbrace{{\bf b},\dots ,{\bf b}}_{\mu},\underbrace{{\bf c},\dots {\bf c}}_{N-\nu-\mu})=\sum_{i_{\alpha},i_{\beta},j_{\alpha},j_{\beta}}|I(i_{\alpha},i_{\beta},j_{\alpha},j_{\beta})| \frac{a_1^{i_{\alpha}}b_1^{i_{\beta}}c_1^{i_{\gamma}}
			a_2^{j_{\alpha}}b_2^{j_{\beta}}c_2^{j_{\gamma}}}{\binom{N}{i,j}}      $$
		where $|I(i_{\alpha},i_{\beta},j_{\alpha},j_{\beta})|$ denotes the cardinality of the set $I(i_{\alpha},i_{\beta},j_{\alpha},j_{\beta})$. 
		We have
		\begin{equation}\label{e16}
			|I(i_{\alpha},i_{\beta},j_{\alpha},j_{\beta})|=
			\binom{\nu}{i_{\alpha}}
			\binom{\mu}{i_{\beta}}
			\binom{N-\nu -\mu}{i_{\gamma}}
			\binom{\nu -i_{\alpha}}{j_{\alpha}}
			\binom{\mu -i_{\beta}}{j_{\beta}}
			\binom{N-\nu-\mu -i_{\gamma}}{j_{\gamma}}.
		\end{equation}
		Hence,
		\begin{eqnarray}	
			&&b_{ij}^{\triangle} (\underbrace{{\bf a},\dots ,{\bf a}}_{\nu},\underbrace{{\bf b},\dots ,{\bf b}}_{\mu},\underbrace{{\bf c},\dots {\bf c}}_{N-\nu-\mu})=\nonumber\\
			&&	\label{e155}\\
			&&	\sum_{j_{\beta}=\underline{j}_{\beta}}^{\bar{j}_{\beta}}
			\sum_{j_{\alpha}=\underline{j}_{\alpha}}^{\bar{j}_{\alpha}}
			\sum_{i_{\beta}=\underline{i}_{\beta}}^{\bar{i}_{\beta}}
			\sum_{i_{\alpha}=\underline{i}_{\alpha}}^{\bar{i}_{\alpha}}
			\binom{\nu}{i_{\alpha}}
			\binom{\mu}{i_{\beta}}
			\binom{\lambda}{i_{\gamma}}
			\binom{\nu -i_{\alpha}}{j_{\alpha}}
			\binom{\mu -i_{\beta}}{j_{\beta}}
			\binom{\lambda -i_{\gamma}}{j_{\gamma}}
			\frac{a_1^{i_{\alpha}}b_1^{i_{\beta}}c_1^{i_{\gamma}}
				a_2^{j_{\alpha}}b_2^{j_{\beta}}c_2^{j_{\gamma}}}{\binom{N}{i,j}},\nonumber
		\end{eqnarray}
		where $\lambda :=N-\nu -\mu$.
		It remains to define $\underline{i}_{\alpha},\bar{i}_{\alpha},\underline{i}_{\beta},\bar{i}_{\beta},\underline{j}_{\alpha},\bar{j}_{\alpha},\underline{j}_{\beta},\bar{j}_{\beta}$. These bounds can be computed from the conditions they satisfy as follows.
		\begin{equation}\label{e10}
			0\leq i_{\alpha}\leq\nu,\ 0\leq i_{\beta}\leq\mu,\ i_{\alpha}+i_{\beta}\leq i,
		\end{equation}
		\begin{equation}\label{e111}
			i_{\gamma}=i-i_{\alpha}-i_{\beta}\leq N-\nu-\mu=\lambda\ \Leftrightarrow\ i-\lambda\leq i_{\alpha}+i_{\beta}
		\end{equation}
		From (\ref{e10}) we have $i_{\alpha}\leq\nu$ and $i_{\alpha}\leq i-i_{\beta}\leq i$. From (\ref{e111}) we have $$i_{\alpha}\geq i-\lambda -i_{\beta}\geq i-\lambda-\mu =i+\nu-N.$$  We obtained
		$\max (0,i+\nu-N)\leq i_{\alpha}\leq \min (i,\nu)$ and hence, $$\bar{i}_{\alpha}=\min (i,\nu),\ \underline{i}_{\alpha}=\max (0,i+\nu-N).$$
		Next, from (\ref{e10}) we have $i_{\beta}\leq \min (i-i_{\alpha},\mu)$. Frow (\ref{e111}) it follows $i_{\beta}\geq \max (0, i-i_{\alpha}-\lambda)$. Hence, $\bar{i}_{\beta}=\min (i-i_{\alpha},\mu)$ and $\underline{i}_{\beta}=\max (0, i-i_{\alpha}-\lambda)$.
		Further, we have
		\begin{equation}\label{e22}
			0\leq i_{\alpha}+j_{\alpha}\leq\nu,\ 0\leq i_{\beta}+j_{\beta}\leq\mu,\ j_{\alpha}+j_{\beta}\leq j.
		\end{equation}
		\begin{equation}\label{e133}
			j_{\gamma}=j-j_{\alpha}-j_{\beta}\leq N-\nu-\mu -(i-i_{\alpha}-i_{\beta})=\lambda -i_{\gamma}.
		\end{equation}
		From (\ref{e22}) it follows $j_{\alpha}\leq\min (j, \nu-i_{\alpha})$. From (\ref{e133}) and (\ref{e22}) we obtain
		$$j_{\alpha}\geq j-j_{\beta}-(\lambda -i_{\gamma})\geq j-(\mu -i_{\beta})-(\lambda -i_{\gamma}).$$
		Hence, $\bar{j}_{\alpha}=\min (j, \nu-i_{\alpha})$,\ $\underline{j}_{\alpha}=j-(\mu -i_{\beta})-(\lambda -i_{\gamma}).$
		Now it remains to find $\bar{j_{\beta}}$ and $\underline{j}_{\beta}$. From (\ref{e22}) we have $$j_{\beta}\leq\min(j-j_{\alpha},\mu-i_{\beta}).$$
		From (\ref{e133}) it follows $j_{\beta}\geq\max (0,j-j_{\alpha}-(\lambda-i_{\gamma}))$. Hence,
		$$\bar{j}_{\beta}=\min(j-j_{\alpha},\mu-i_{\beta}),\ \underline{j}_{\beta}=\max (0,j-j_{\alpha}-(\lambda-i_{\gamma})).$$
		Finally, we obtain
		\begin{eqnarray*}	
			{\bf q}_{\nu\mu}&=&\sum_{i=0}^n\sum_{j=0}^m\frac{{\bf c}_{ij}}{\binom{N}{i,j}}		
			\sum_{i_{\alpha}=\underline{i}_{\alpha}}^{\bar{i}_{\alpha}}	
			\sum_{i_{\beta}=\underline{i}_{\beta}}^{\bar{i}_{\beta}}	
			\sum_{j_{\alpha}=\underline{j}_{\alpha}}^{\bar{j}_{\alpha}}
			\sum_{j_{\beta}=\underline{j}_{\beta}}^{\bar{j}_{\beta}}\\
			&&	\binom{\nu}{i_{\alpha}}
			\binom{\mu}{i_{\beta}}
			\binom{\lambda}{i_{\gamma}}
			\binom{\nu -i_{\alpha}}{j_{\alpha}}
			\binom{\mu -i_{\beta}}{j_{\beta}}
			\binom{\lambda -i_{\gamma}}{j_{\gamma}}
			a_1^{i_{\alpha}}b_1^{i_{\beta}}c_1^{i_{\gamma}}
			a_2^{j_{\alpha}}b_2^{j_{\beta}}c_2^{j_{\gamma}},\\
			&&  \nu=0,\dots ,N,\ \mu=0,\dots ,N-\nu ,
		\end{eqnarray*}
		
		where
		$$
		\begin{array}{l}
			N=n+m,\ \lambda=N-\nu -\mu ,\  i_{\gamma}=i-i_{\alpha}-i_{\beta},\ j_{\gamma}=j-j_{\alpha}-j_{\beta},\\
			\underline{i}_{\alpha}=\max (0,i+\nu -N),\ \bar{i}_{\alpha}=\min (i,\nu),\\
			\underline{i}_{\beta}=\max (0,i-i_{\alpha}-\lambda),\ \bar{i}_{\beta}=\min (i-i_{\alpha},\mu),\\
			\underline{j}_{\alpha}=\max (0,j-(\mu-i_{\beta})-(\lambda - i_{\gamma})),\  \bar{j}_{\alpha}=\min (j,\nu -i_{\alpha}),\\
			\underline{j}_{\beta}=\max (0,j-j_{\alpha}-(\lambda - i_{\gamma})),\  \bar{j}_{\beta}=\min (j-j_{\alpha},\mu -i_{\beta}).
		\end{array}
		$$
\hfill$\Box$
	\begin{Remark}\label{rem1}
		The cardinality of the set $I(i_{\alpha},i_{\beta},j_{\alpha},j_{\beta})$ in
		(\ref{e155}) can also be presented in an equivalent way as follows.
		\begin{equation}\label{e17}
			|I(i_{\alpha},i_{\beta},j_{\alpha},j_{\beta})|=
			\binom{\nu}{j_{\alpha}}
			\binom{\mu}{j_{\beta}}
			\binom{N-\nu -\mu}{j_{\gamma}}
			\binom{\nu -j_{\alpha}}{i_{\alpha}}
			\binom{\mu -j_{\beta}}{i_{\beta}}
			\binom{N-\nu-\mu -j_{\gamma}}{j_{\gamma}}
		\end{equation}
		
	\end{Remark}
\section{Examples and Results}
\label{sect3}

We have implemented and tested the obtained formulae  using the Mathematica package \citep{Mathematica}. In this section we present the results from our experimental work.

\begin{Example}\label{example1}
	
	{\rm We consider the parametric polynomial surface $S(u,v)=\left(x(u,v),y(u,v),z(u,v)\right)$  where the coordinate functions $x(u,v),y(u,v),z(u,v)$ are as follows:
		$$
		\begin{array}{lcl}
			x(u,v)&=&3 u + u^3,\\[1ex]
			y(u,v)&=&\frac{7}{5} v + \frac{9}{5} u v - \frac{9}{5} u^2 v + \frac{13}{5} u^3 v + \frac{3}{5} v^2 - \frac{24}{5} u v^2 + \frac{39}{5} u^2 v^2 - \frac{23}{5} u^3 v^2\\[1ex]
			z(u,v)&=&3 u - \frac{15}{4} u^2 + \frac{19}{20} u^3 - v - 3 u v + \frac{117}{10} u^2 v - \frac{38}{5} u^3 v + \frac{1}{5}v^2 + \frac{27}{5} u v^2\\[1ex]
			&&- \frac{141}{10} u^2 v^2 + \frac{179}{20} u^3 v^2.
		\end{array}	
		$$
		
		The surface $S$ is represented as $S(u,v)=\sum_{i=0}^n\sum_{j=0}^m{\bf c}_{ij}u^iv^j$ where $n=3$ and $m=2$. The corresponding coefficients ${\bf c}_{ij}$, $i=0,\dots ,3$, $j=0,\dots ,2$, are shown in Table~\ref{table1}. We applied formula (\ref{eq1}) for $a=0$, $b=1$, $c=0$, $d=1$ and obtained TPB surface $S_1$ of degree $(3,2)$. The control points of $S_1$ are shown in Table~\ref{table1}. The surface $S_1$ and its control polygon are shown in Figure~\ref{fig1}{\rm ({\bf a})}.
		
		Further, we applied formula (\ref{eq1}) for $a=1/3$, $b=2/3$, $c=1/4$, $d=3/4$ and obtained the TPB surface $S_2$ of degree $(3,2)$. The control points of $S_2$ are shown in Table~\ref{table1}. Both surfaces $S_1$ and $S_2$ with their control polygons are shown in Figure~\ref{fig1}{\rm ({\bf b})}.
	}
	
	\begin{figure}[htbp]
				\centering
			\subfloat[\centering]{\includegraphics[width=6.5cm]{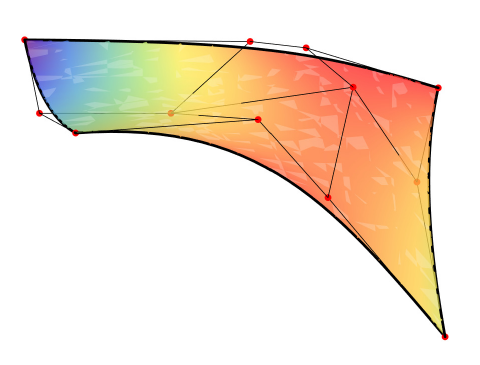}}
						\subfloat[\centering]{\includegraphics[width=6.5cm]{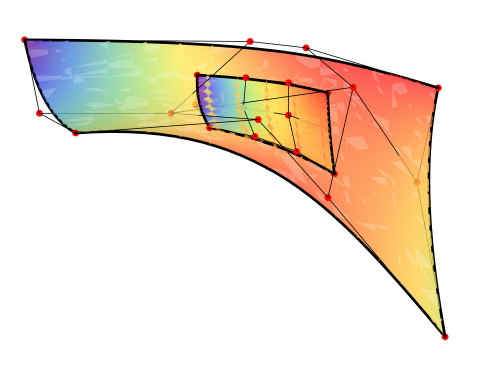}}
				\caption{Subdivision of TPB surfaces Example~\ref{example1} using formula (\ref{eq1}): (\textbf{a}) The TPB surface $S_1(u,v)$ defined for $0\leq u\leq 1$, $0\leq v\leq 1$. (\textbf{b}) $S_1$ and the TPB surface $S_2(u,v)$ defined for $\frac{1}{3}\leq u\leq \frac{2}{3}$, $\frac{1}{4}\leq v\leq \frac{3}{4}$.\label{fig1}}
	\end{figure}

	\begin{table}[htbp]
		\caption{ The coefficients of surface $S$ and the control points of TPB surfaces $S_1$ and $S_2$ from Example~\ref{example1}\label{table1}}
								\centering
				 	\begin{tabular}{llll}
					\hline
					\noalign{\smallskip}
						{ $S(u,v)$}
						& ${\bf c}_{00}=(0, 0, 0)$			& ${\bf c}_{01}=(0, \frac{7}{5}, -1)$			& ${\bf c}_{02}=(0, \frac{3}{5}, \frac{1}{5})$\\[1ex]
						& ${\bf c}_{10}=(3,0,3)$			& ${\bf c}_{11}=(0,\frac{9}{5},-3)$			& ${\bf c}_{12}=(0,-\frac{24}{5},\frac{27}{5})$\\[1ex]
						& ${\bf c}_{20}=(0,0,-\frac{15}{4})$			& ${\bf c}_{21}=(0,-\frac{9}{5},\frac{117}{10})$			& ${\bf c}_{22}=(0,\frac{39}{5},-\frac{141}{10})$\\[1ex]
						&  ${\bf c}_{30}=(1,0,\frac{19}{20})$           & ${\bf c}_{31}=(0,\frac{13}{5},-\frac{38}{5})$           &   ${\bf c}_{32}=(0,-\frac{23}{5},\frac{179}{20})$               \\[1ex]
						\noalign{\smallskip}\hline\noalign{\smallskip}
						{$S_1(u,v),$}    &$ {\bf p}_{00}=(0, 0, 0)$ &$ {\bf p}_{01}=(0, \frac{7}{10}, -\frac{1}{2})$  &$ {\bf p}_{02}=(0, 2, -\frac{4}{5})$  \\[1ex]
						& ${\bf p}_{10}=(1, 0, 1)$& ${\bf p}_{11}=(1, 1, 0)$ & ${\bf p}_{12}=(1, 1, 1)$  \\[1ex]
						$0\leq u\leq 1$	& ${\bf p}_{20}=(2, 0, \frac{3}{4})$ &$ {\bf p}_{21}=(2, 1, \frac{6}{5})$ &$ {\bf p}_{22}=(2, 2, \frac{3}{4})$ \\[1ex]
						$0\leq v\leq 1$	& ${\bf p}_{30}=(4, 0, \frac{1}{5})$ & ${\bf p}_{31}=(4, 2, \frac{1}{4})$ & ${\bf p}_{32}=(4, 3, \frac{3}{4})$\\[1ex]
						\noalign{\smallskip}\hline\noalign{\smallskip}
						{$S_2(u,v),$}     &$ {\bf p}_{00}=(\frac{28}{27}, \frac{983}{2160}, \frac{3637}{8640})$ &$ {\bf p}_{01}=(\frac{28}{27}, \frac{385}{432}, \frac{781}{2880})$  &$ {\bf p}_{02}=(\frac{28}{27}, \frac{901}{720}, \frac{2701}{8640})$  \\[1ex]
						& ${\bf p}_{10}=(\frac{38}{27}, \frac{527}{1080}, \frac{613}{1080})$& ${\bf p}_{11}=(\frac{38}{27}, \frac{205}{216}, \frac{7}{15})$ & ${\bf p}_{12}=(\frac{38}{27}, \frac{469}{360}, \frac{571}{1080})$  \\[1ex]
						$\frac{1}{3}\leq u\leq \frac{2}{3}$		& ${\bf p}_{20}=(\frac{49}{27}, \frac{1157}{2160}, \frac{275}{432})$ &$ {\bf p}_{21}=(\frac{49}{27}, \frac{451}{432}, \frac{431}{720})$ &$ {\bf p}_{22}=(\frac{49}{27}, \frac{1039}{720}, \frac{1399}{2160})$ \\[1ex]
						$\frac{1}{4}\leq v\leq \frac{3}{4}$	& ${\bf p}_{30}=(\frac{62}{27}, \frac{1321}{2160}, \frac{265}{432})$ & ${\bf p}_{31}=(\frac{62}{27}, \frac{515}{432}, \frac{449}{720})$ & ${\bf p}_{32}=(\frac{62}{27}, \frac{1187}{720}, \frac{1469}{2160})$\\[1ex]
						\noalign{\smallskip}\hline\noalign{\smallskip}
					\end{tabular}
								\end{table}

		\end{Example}

		\begin{Example}\label{example2}
			{\rm We consider the parametric polynomial surface $S$ defined in Example~\ref{example1}. First, we applied formula (\ref{eq2}) for
				the triangle $\triangle \bf{abc}$ with vertices ${\bf a}=(0,0)$, ${\bf b}=(1,0)$, ${\bf c}=(0,1)$		
				and obtained the TB surface $S_3$ of total degree $5$. The 21 control points of $S_3$ are shown in Table~\ref{table2}. The surface $S_3$ and its control polygon are shown in Figure~\ref{fig2}{\rm ({\bf a})}. The TPB surface $S_1$ and the TB surface $S_3$ with its control polygon are shown in Figure~\ref{fig2}{\rm ({\bf b})}.
				
				Further, we applied formula (\ref{eq2}) for the triangle with vertices ${\bf a}=(0,\frac{1}{2})$, ${\bf b}=(\frac{1}{2},0)$, ${\bf c}=(\frac{1}{2},\frac{1}{2})$ and obtained the TPB surface $S_4$ of degree $5$. The 21 control points of $S_4$ are shown in Table~\ref{table2}. The TB surface $S_4$ and its control polygon are shown in Figure~\ref{fig2}{\rm ({\bf c})}. Both surfaces $S_3$ and $S_4$ with its control polygon are shown in Figure~\ref{fig2}{\rm ({\bf d})}.
			}

		\begin{figure}[htbp]
			
			\centering
			\subfloat[\centering]{\includegraphics[width=6.cm]{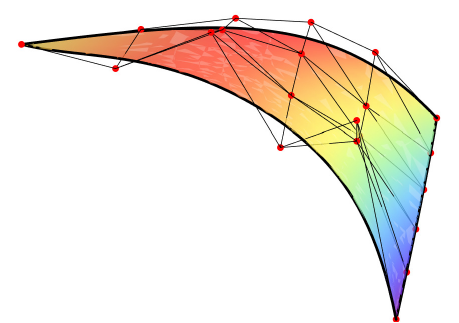}}
			\subfloat[\centering]{\includegraphics[width=6.5cm]{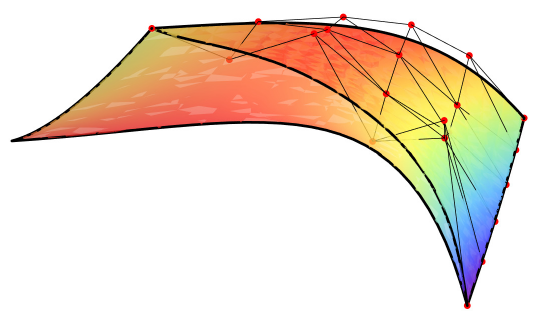}}\\
			
			\subfloat[\centering]{\includegraphics[width=6.5cm]{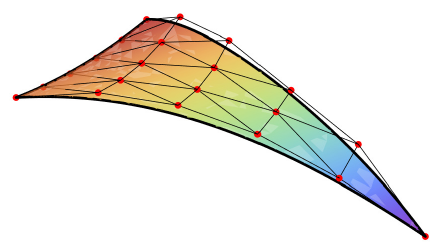}}
			\subfloat[\centering]{\includegraphics[width=6.cm]{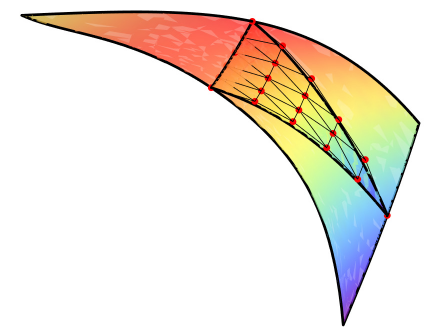}}
			
			\caption{Subdivision of TB surfaces from Example~\ref{example2} using formula (\ref{eq2}): (\textbf{a}) The TB surface $S_3$ of total degree 5 defined in the triangle with vertices ${\bf a}=(0,0)$, ${\bf b}=(1,0)$, ${\bf c}=(0,1)$. (\textbf{b}) The TPB surface $S_1$ and the TB surface $S_3$ with its control polygon. (\textbf{c}) The TB surface $S_4$ of total degree 5 defined in the triangle with vertices ${\bf a}=(0,\frac{1}{2})$, ${\bf b}=(\frac{1}{2},0)$, ${\bf c}=(\frac{1}{2},\frac{1}{2})$ and its control polygon. (\textbf{d}) The TB surfaces $S_3$ and $S_4$ with its control polygon. \label{fig2}}
		\end{figure}

			\begin{table}[htbp]
				\caption{ The control points of TB surfaces $S_3$ and $S_4$ of degree $5$ from Example~\ref{example2}\label{table2}}
												\centering
								\begin{tabular}{cllll}
								\hline
								\noalign{\smallskip}
								{ $S_3$}
								& ${\bf q}_{05}=(0, 0, 0)$  &${\bf q}_{04}=(\frac{3}{5}, 0, \frac{3}{5})$ &${\bf q}_{14}=(0, \frac{7}{25}, -\frac{1}{5})$\\[1ex]
								&${\bf q}_{03}=(\frac{6}{5}, 0, \frac{33}{40})$
								&${\bf q}_{13}=(\frac{3}{5}, \frac{37}{100}, \frac{1}{4})$ &${\bf q}_{23}=(0, \frac{31}{50}, -\frac{19}{50})$\\[1ex]
							${\bf a}=(0,0)$	&${\bf q}_{02}=(\frac{19}{10}, 0, \frac{77}{100})$
								&${\bf q}_{12}=(\frac{6}{5}, \frac{2}{5}, \frac{143}{200})$ &${\bf q}_{22}=(\frac{3}{5}, \frac{16}{25}, \frac{1}{10})$\\[1ex]
							${\bf b}=(1,0)$	&${\bf q}_{32}=(0, \frac{51}{50}, -\frac{27}{50})$ &${\bf q}_{01}=(\frac{14}{5}, 0, \frac{53}{100})$
								&${\bf q}_{11}=(\frac{19}{10}, \frac{1}{2}, \frac{91}{100})$\\[1ex]
							${\bf c}=(0,1)$	&${\bf q}_{21}=(\frac{6}{5}, \frac{4}{5}, \frac{103}{200})$ &${\bf q}_{31}=(\frac{3}{5}, \frac{81}{100}, \frac{3}{20})$ &${\bf q}_{41}=(0, \frac{37}{25}, -\frac{17}{25})$\\[1ex]
								&${\bf q}_{00}=(4, 0, \frac{1}{5})$
								&${\bf q}_{10}=(\frac{14}{5}, \frac{4}{5}, \frac{11}{20})$ &${\bf q}_{20}=(\frac{19}{10}, \frac{9}{10}, \frac{219}{200})$\\[1ex]
								&${\bf q}_{30}=(\frac{6}{5}, \frac{6}{5}, \frac{9}{40})$ &${\bf q}_{40}=(\frac{3}{5}, \frac{22}{25}, \frac{2}{5})$ &${\bf q}_{50}=(0, 2, -\frac{4}{5})$\\[1ex]						
								\noalign{\smallskip}\hline\noalign{\smallskip}
								{$S_4$}
								& ${\bf q}_{05}=(0, \frac{17}{20}, -\frac{9}{20})$  &${\bf q}_{04}=(\frac{3}{10}, \frac{31}{50}, -\frac{17}{200})$ &${\bf q}_{14}=(\frac{3}{10}, \frac{41}{50}, -\frac{33}{200})$\\[1ex]
								&${\bf q}_{03}=(\frac{3}{5}, \frac{81}{160}, \frac{303}{1600})$
								&${\bf q}_{13}=(\frac{3}{5}, \frac{523}{800}, \frac{43}{320})$ &${\bf q}_{23}=(\frac{3}{5}, \frac{653}{800}, \frac{27}{320})$\\[1ex]
								${\bf a}=(0,\frac{1}{2})$&${\bf q}_{02}=(\frac{73}{80}, \frac{601}{1600}, \frac{2963}{6400})$
								&${\bf q}_{12}=(\frac{73}{80}, \frac{857}{1600}, \frac{2419}{6400})$ &${\bf q}_{22}=(\frac{73}{80}, \frac{221}{320}, \frac{2051}{6400})$\\[1ex]
							${\bf b}=(\frac{1}{2},0)$	&${\bf q}_{32}=(\frac{73}{80}, \frac{269}{320}, \frac{1859}{6400})$ &${\bf q}_{01}=(\frac{5}{4}, \frac{87}{400}, \frac{253}{400})$
								&${\bf q}_{11}=(\frac{5}{4}, \frac{2}{5}, \frac{3551}{6400})$\\[1ex]
							${\bf c}=(\frac{1}{2},\frac{1}{2})$	&${\bf q}_{21}=(\frac{5}{4}, \frac{459}{800}, \frac{1593}{3200})$ &${\bf q}_{31}=(\frac{5}{4}, \frac{591}{800}, \frac{2953}{6400})$ &${\bf q}_{41}=(\frac{5}{4}, \frac{179}{200}, \frac{713}{1600})$\\[1ex]
								&${\bf q}_{00}=(\frac{13}{8}, 0, \frac{109}{160})$
								&${\bf q}_{10}=(\frac{13}{8}, \frac{87}{400}, \frac{503}{800})$ &${\bf q}_{20}=(\frac{13}{8}, \frac{679}{1600}, \frac{3767}{6400})$\\[1ex]
								&${\bf q}_{30}=(\frac{13}{8}, \frac{993}{1600}, \frac{3589}{6400})$ &${\bf q}_{40}=(\frac{13}{8}, \frac{129}{160}, \frac{349}{640})$ &${\bf q}_{50}=(\frac{13}{8}, \frac{157}{160}, \frac{347}{640})$\\[1ex]
							\noalign{\smallskip}\hline\noalign{\smallskip}
							\end{tabular}
												\end{table}

				\end{Example}

				\section{Conclusions}\label{sect4}
				
				Blossoming is an important tool in the analysis and manipulation of polynomial and piecewise polynomial curves and surfaces, significantly enhancing the capabilities within theoretical and applied CAGD . This paper addresses the challenge of deriving control points for subdivided B\'{e}zier curves and surface patches via blossoming. The proposed 
				closed-form formulae computationally optimize the  evaluation process for subdivision of B\'{e}zier curves, and TPB and TB surfaces.
				We believe that the presented evaluation process based on closed-form formulae can be significantly optimized further by providing a more efficient computational procedure 
				computing control points in bulk instead of computing each control point separately. Another direction of research is to consider blossoming approach in non-polynomial cases and 
				look for speed-up computational procedures for evaluation of the corresponding blossoms.

\section*{Acknowledgments}

This research was supported in part by Sofia University Science Fund Grant No. 80-10-11/2025 and the project UNITe BG16RFPR002-1.014-0004 funded by PRIDST.

\bibliographystyle{elsarticle-harv}\biboptions{authoryear}

\bibliography{arxiv2025bibfile}

\end{document}